\documentclass{article}
\pdfoutput=1
\pdfoutput=1 
\usepackage[utf8]{inputenc}
\usepackage{caption}
\usepackage{subcaption}
\usepackage{multicol}
\usepackage{xcolor}
\usepackage{fancyhdr}
\usepackage{cite}
\usepackage{graphics}
\usepackage{wrapfig}
\usepackage{float}
\usepackage{hyperref}
\usepackage{amsmath}
\hypersetup{
    colorlinks,
    citecolor=blue,
    filecolor=blue,
    linkcolor=blue,
    urlcolor=blue
}

\usepackage[bottom]{footmisc}
\usepackage{titling}
\usepackage{ulem}
\predate{}
\postdate{}

\usepackage{lipsum}
\usepackage{graphicx}
\usepackage[T1]{fontenc}

\newcommand{\ppt}{$p_{\rm T}$}

\newcommand{\pythia}{\textsc{Pythia}}

\def \pp    {$pp$ }

\date{} 
\begin{document}


\begin{center}
{\bf Testing of \pythia{} modes to study identified particle production in high-multiplicity $\mathbf{pp}$ collisions at $\mathbf{\sqrt{s}}$ = 7\;TeV }

\vskip1.0cm
Rabia Bashir$^{1}$,
Ramoona Shehzadi$^{1}$ and 
M.~U.~Ashraf$^{2}$ {\footnote{usman.ashraf@cern.ch;}}

{\small\it

$^1$ Department of Physics, University of the Punjab, Lahore 54590, Pakistan\\
$^2$ Centre for Cosmology, Particle Physics and Phenomenology (CP3), Université Catholique de Louvain, B-1348 Louvain-la-Neuve, Belgium\\

}


\begin{abstract}

This study presents a comprehensive analysis of particle production in proton-proton (\pp) collisions at $\sqrt{s}$ = 7 TeV using \pythia~8 event generator. We investigate the transverse momentum \ppt spectra of light charged hadrons ($\pi^\pm$, $K^\pm$ and $p(\bar p)$), their yield ratios ($\pi^-/\pi^+$, $K^-/K^+$ and $\bar{p}/p$), and \ppt-differential ratios ($(K^++K^-)/(\pi^++\pi^-)$, $(\overline{p}+p)/(\pi^++\pi^-)$) and mean transverse momentum ($\langle p_\mathrm{T} \rangle$). Our analysis employs various \pythia~8 tunes (Simple, Vincia, and Dire) to explore the impact of different model configurations on particle production. We optimize a key parameter ($p_\mathrm{T}HatMin$) within each tune to achieve the best agreement between the simulated \ppt spectra and those measured by the CMS collaboration. Interestingly, we find that the optimal values for $p_\mathrm{T}HatMin$ differ between hadron species, potentially reflecting the influence of particle mass on production mechanisms. 
It is not possible to simultaneously and qualitatively describe both, the strangeness enhancement and collectivity in $pp$ collisions from \pythia~8. Further investigation such as final-state effects such as color ropes or junctions may require to explain these effects. These types of studies help us identify limitations in current models and refine their parameters to better explain experimental observations.

\vskip0.5cm

\end{abstract}
\end{center}

\begin{multicols}{2}
\section{Introduction}\label{sec1}

Quantum Chromodynamics (QCD) suggests a phase transition from a high energy density and high temperature known as the Quark Gluon Plasma (QGP), a deconfined state of quarks and gluons, to a normal hadronic matter where the relevant degrees of freedom are more hadronic~\cite{1, 2, 3, 4, 5}. The clear indication of the formation of a QGP-like new state of matter has been reported by several experiments at the Relativistic Heavy Ion Collider (RHIC) and the Large Hadron Collider (LHC)~\cite{6, 7, 8, 9}. The ``fire-ball'' created in the heavy-ion collisions is a fluid of strongly interacting matter, the QGP or sQGP which has been primarily characterized by the collective flow of the final state particles. The various experimental observables of QGP related studies have been reported by many collaborations~\cite{10, 11, 12, 13, 14, 15, 16, 17, 18, 19, 20, 21, 22}. The data from proton-proton ($pp$) collisions at the same center of mass energy may serve as a baseline to study the signatures of the QGP in heavy-ion collisions. This is due to the fact that the hydrodynamic system produced in $pp$ collisions is expected to be different from that produced in heavy-ion collisions. The hydrodynamic system produced in heavy-ion collisions is according to our understanding based on phenomenological and theoretical models. On the other hand, system produced in $pp$ collisions needs further investigation.  

Measurements of hadron production hold immense significance in the domains of cosmic-ray, nuclear, and high-energy particle physics. At collider energies, the hadrons at the partonic level are mainly produced by soft and hard scattering processes. In hard scatterings, large momentum transfer between two partons results the production of high {\ppt} particles, while majority of low {\ppt} ({\ppt} $< 2$ GeV/$c$) particles are produced due to soft scattering processes, where small amount of momentum is transferred between two partons~\cite{23, 24, 25}. QCD stands as the successful theoretical framework for describing strong interactions and the diverse characteristics of particle spectra in high-energy interactions. In $pp$ collisions, the production of hadrons at high {\ppt} finds its explanation within perturbative QCD (pQCD). This process involves hard parton-parton scattering followed by fragmentation~\cite{23, 26}. Bulk observables, including particle spectra, charged-particle multiplicities, and particle ratios, provide an excellent probe for investigating the properties of the QGP~\cite{26a}. The {\ppt} spectra of hadrons play a crucial role in unraveling the interactions of soft partons and the process of hadronization, as envisioned by non-perturbative QCD. However, for small average momentum transfers, direct evaluation from first principles within pQCD is not feasible. The $p/\pi $ ratio serves as an indicator of the relative production of baryons compared to meson and $K/\pi$ ratio serve as a measure of strangeness enhancement. The correlation between these observables provide a comprehensive understanding of the interplay between soft and hard processes in the collisions, and thus shed light on the equation of state of the hot hadronic matter~\cite{26a}. Predictions in these scenarios rely on phenomenological models, given the complexities of precise calculations in non-perturbative processes. Monte Carlo generators built upon these models require tuning and optimization to replicate measurable quantities observed in experiments. Additionally, these models encompass multi-parton processes, making it challenging to ascertain their energy evolution. By comparing identified particles in the low {\ppt} regime with QCD-inspired models, it becomes possible to establish a reference for comparisons at higher energies.

This article presents the results on {\ppt} spectra of identified light charged hadrons, the antiparticle to particle ratios of same hadrons under study, the ratios of different hadrons such as $(K^\pm)/(\pi^\pm)$ and $(\overline{p}+p)/(\pi^++\pi^-)$ and mean transverse momentum ($\langle p_\mathrm{T} \rangle$) in $pp$ collisions at $\sqrt{s}$= 7 TeV from \pythia~8 with different tunes (Simple, Vincia, and Dire showers), with the aim to understand the anomalous features of particle production mechanisms in $pp$ collisions at the LHC. We also tune and propose the optimal values for $p_\mathrm{T}HatMin$ differ between hadron species to study its impact on the bulk observables of these hadrons. These results are then compared with the observations from the CMS experiment~\cite{27}.
 
The paper is organized as follows. The details of \pythia~8 is described in Section~\ref{sec2}. The results and discussion are presented in Section~\ref{sec3}. Finally, the results are summarized in Section~\ref{sec4}.

\section{Event Generator}{\label{sec2}}

In this section, we briefly discuss the pQCD-inspired event generator \pythia~8, its different tunes, and the analysis methedology.

\pythia~8.3 \cite{31} is a multi-purpose Monte-Carlo event generator designed for generating events in particle collisions at higher energies. The fundamental component of \pythia~ historically depends on the Lund string model, which is particularly suitable for hadronization processes. It focuses on the impacts of the strong nuclear interaction controlled by QCD. The software primarily utilizes C++ and seamlessly integrates an extensive collection of intricate physics models to simulate the transition from a few-body hard-scattering event to a complex multi-particle end state. These physics models include theoretically derived components and phenomenological models that rely on data-driven parameter determinations. Additionally, it finds application in various studies within astro, nuclear, neutrino, and particle physics. It is used for studies of an experiment's implications of theoretical hypothesis, interpreting data from experiments, developing search strategies, as well as detector design and performance studies.

\textbf{Showers in \pythia~8.3}: \pythia~8.3 offers three distinct modules of shower namely Simple, Vincia and Dire.
As well as being the standard shower module in \pythia~8.3, the \textbf{Simple shower} is the oldest parton shower algorithm in \pythia~8. The origins of this algorithm can be traced back to the mass-ordered shower employed in JETSET/\pythia~ \cite{32,33,34,35} and their development towards {\ppt} ordering. {\ppt} ordering was motivated through a combination of factors, including the influence of the Lund dipole concept and the objective of integrating the progression of initial–state radiation (ISR) and final–state radiation (FSR) showers to multiple parton interactions (MPI) into a unified interspersed order. \textbf{Vincia} module incorporates a sequential {\ppt} ordered development by the antenna formalist approaches\cite{36,37}. Within the context of event generators, the ARIADNE model, which was extensively employed at The Large Electron Positron Collider, played a pioneering role in introducing this particular type of shower. Vincia exhibits numerous similarities to ARIADNE, particularly in case of FSR QCD radiation. However, in the case of ISR, Vincia treatment differs significantly from ARIADNE. Currently, ARIADNE-style ``global'' showers are still accessible as an alternative option, although not selected by default. Another alternative showering model is the \textbf{Dire} \cite{38}. The aim is to combine the characteristics of parton showers with a dipole in antenna showers. The objective of the hybrid model aims to incorporate a modeling of soft emission impacts from dipole showers, while also maintaining a precise connection between fragmentations and particular coalligned orientations. It facilitates a straightforward position with components found within QCD factorization theorems. 


\section{Results and Discussion}\label{sec3}

In this section, we present results on observables such as {\ppt} spectra, anti-particle to particle ratios ($\pi^-/\pi^+$, $K^-/K^+$ and $\bar{p}/p$), \ppt-differential particle ratios ($(K^++K^-)/(\pi^++\pi^-)$ and $(\overline{p}+p)/(\pi^++\pi^-)$ and mean transverse momentum ($\langle p_\mathrm{T} \rangle$) in $pp$ collisions at $\sqrt{s}$ = 7 TeV using various tunes of \pythia~8 (Simple, Vincia, and Dire).  

\subsection{Transverse momentum {\ppt} spectra}
 
\begin{figure*}[!ht]{\label{fig1}} 
    \centering
    \includegraphics[width=0.75\textwidth]{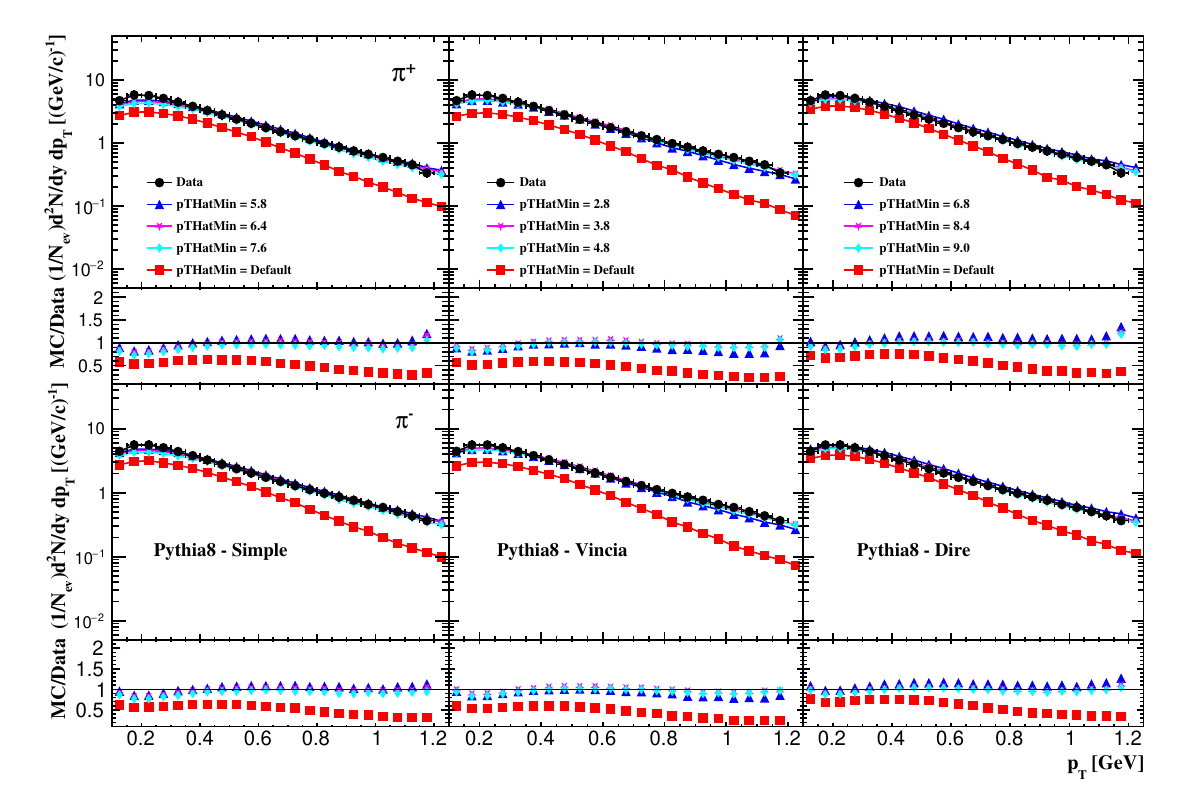}
    \caption{Transverse momentum {\ppt} spectra of $\pi^\pm$ in $pp$ collisions at $\sqrt{s}$= 7\;TeV from \pythia~8 Simple (left), Vincia (middle) and Dire (right). The experimental data for comparison are taken from Ref.\cite{27}. The Model/data ratio is shown in the bottom panel.}
    \label{fig1}
\end{figure*}

Figure~\ref{fig1} shows the transverse momentum {\ppt} spectra of $\pi^\pm$ in $pp$ collisions at $\sqrt{s}$= 7 TeV from various \pythia~8 modes (Simple, Vincia and Dire). These distributions are calculated within the kinematic range defined as $|y| < 1$. The results are then compared with the CMS experimental data~\cite{27}. In \pythia~8, one can turn on more than one processes at the same time. Enabling the $HardQCD:all$ switch turns on all QCD $2 \rightarrow 2$ quark/gluon production processes. Since these processes are divergent at {\ppt} $\rightarrow 0$, therefore, it is necessary to introduce a lower {\ppt} cut-off phase space parameter $p_\mathrm{T}HatMin$ addressing potential issues of instability at exceedingly low $p_{T}$ values.To ensure precise predictions using the \pythia~8, we performed fine-tuning on the $p_\mathrm{T}HatMin$ variable for each module. We varied the values of $p_\mathrm{T}HatMin$ parameter from 0 up to 20\;GeV/$c$ for each particle specie and selected only those values where the {\ppt} spectrum of the pions from different modes of \pythia~8 is in good agreement with the experimental data. The simple mode of \pythia~8, where the default value of $p_\mathrm{T}HatMin$ is 0\;GeV/$c$, does not describe the experimental data, exhibiting a deviation of around 50\%, as evident from the bottom panel of Fig.~\ref{fig1}. The other modes in \pythia~ (Dire and Vincia), also underestimate the experimental data at the default value of $p_\mathrm{T}HatMin$. On the other hand, all three modes of Pythia, with tuned values of the parameter $p_\mathrm{T}HatMin$, reasonably well describe the experimental data. However, please note that the values of $p_\mathrm{T}HatMin$ differ for each mode of \pythia~8 as well as for each particle specie and are listed in Table~\ref{table1}. 

\begin{figure*}[!ht]{\label{fig2}} 
    \centering
    \includegraphics[width=0.75\textwidth]{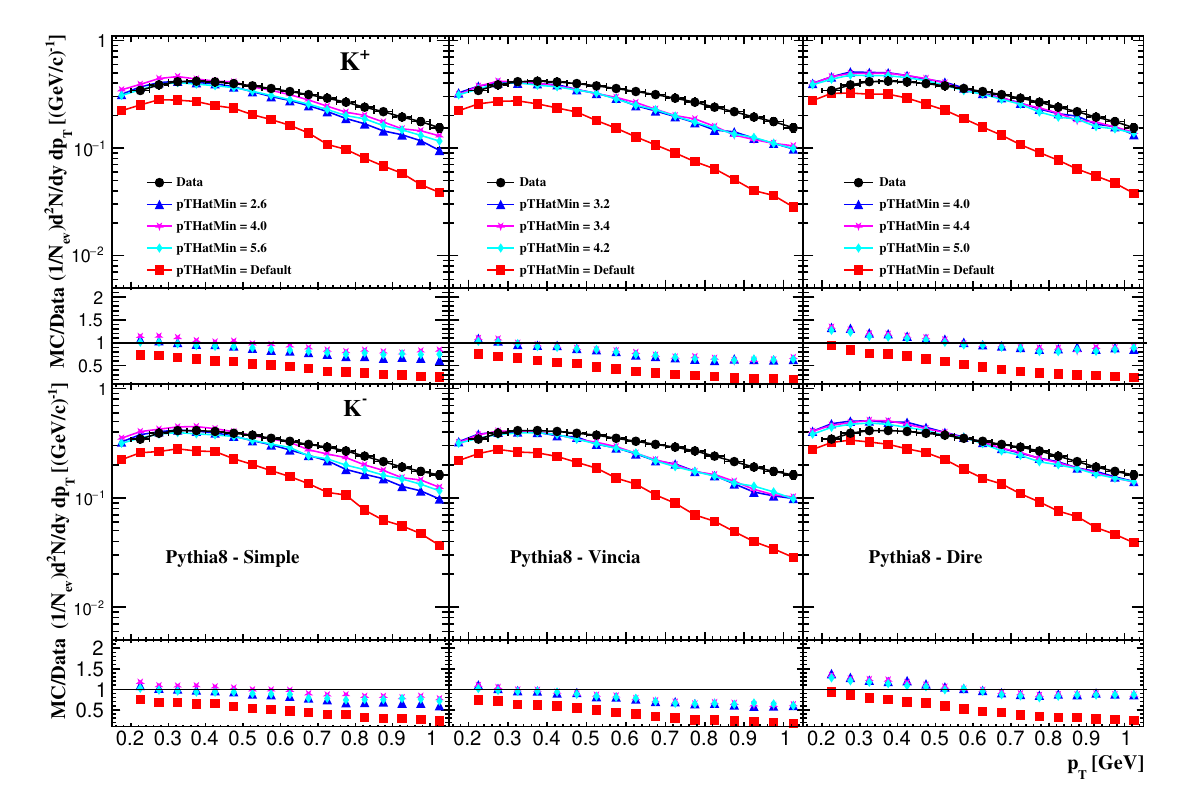}
    \caption{Transverse momentum {\ppt} spectra of $K^\pm$ in $pp$ collisions at $\sqrt{s}$= 7 TeV from \pythia~8 Simple (left), Vincia (middle) and Dire (right). The experimental data for comparison are taken from Ref.\cite{27}. The Model/data ratio is shown in the bottom panel.}
    \label{fig2}
\end{figure*}
Figure~\ref{fig2} illustrates the normalized distributions of transverse momentum {\ppt} spectra of $K^\pm$ in $pp$ collisions at $\sqrt{s}$= 7\;TeV from various \pythia~8 modes (Simple, Vincia and Dire). \pythia~8 parton shower models exhibited varying degrees of accuracy in reproducing the \ppt spectra of $K^\pm$ across the given range. Notably, all three \pythia~8 tunes (Simple, Vincia, and Dire) with the default $p_\mathrm{T}HatMin$ value failed to describe the experimental data, systematically underestimating the yields of $K^\pm$ at all measured \ppt range. In contrast, the tuned values of $p_\mathrm{T}HatMin$ provides better agreement. The tuned values of $p_\mathrm{T}HatMin$ in \pythia~8 Dire gives good agreement with data at {\ppt} $> 0.5$\;GeV/$c$, while \pythia~8 Simple and Vincia agree with data at {\ppt} $< 0.5$\;GeV/$c$. The bottom panel of fig.\ref{fig2} shows the deviation of \pythia~ results with respect to the data. The optimal values for $p_\mathrm{T}HatMin$ for $K^\pm$ are listed in Table~\ref{table1}.

\begin{figure*}[!ht]{\label{fig3}} 
    \centering
    \includegraphics[width=0.75\textwidth]{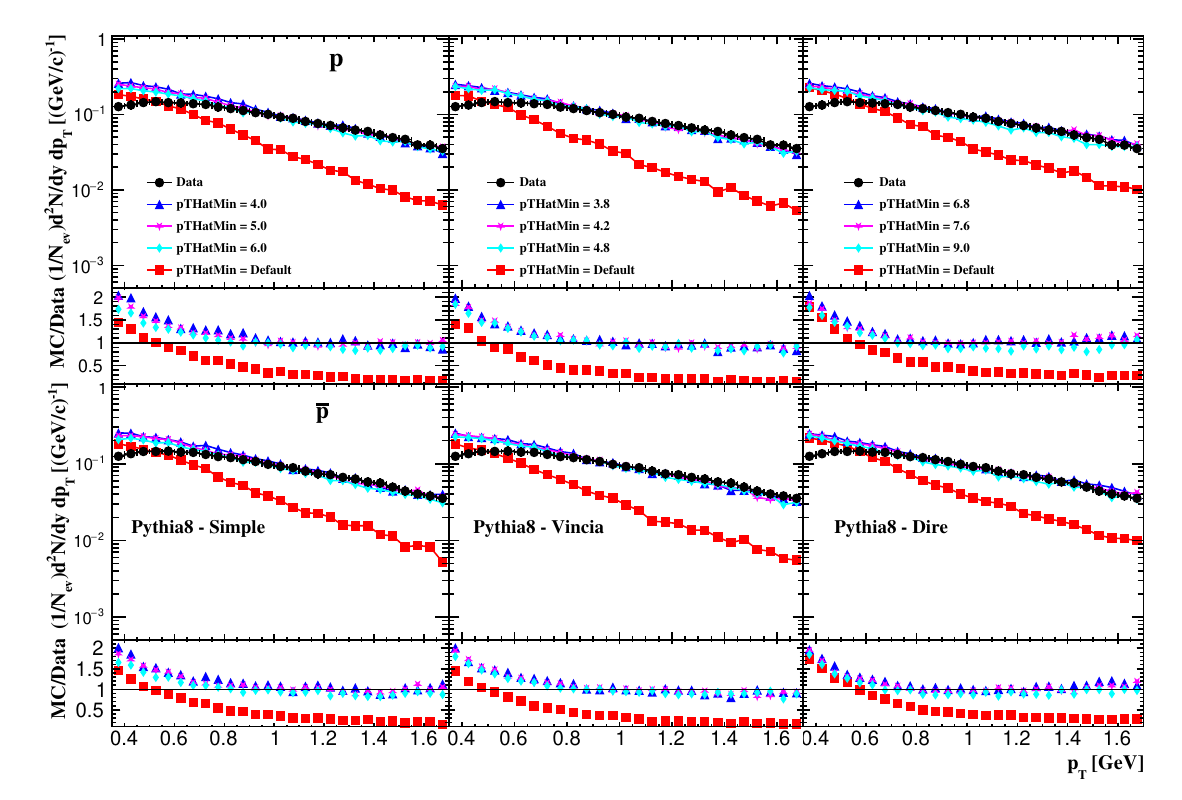}
    \caption{Transverse momentum {\ppt} spectra of $p (\overline{p})$ in $pp$ collisions at $\sqrt{s}$= 7 TeV from \pythia~8 Simple (left), Vincia (middle) and Dire (right). The experimental data for comparison are taken from Ref.\cite{27}. The Model/data ratio is shown in the bottom panel.}
    \label{fig3}
\end{figure*}

Figure~\ref{fig3} shows the {\ppt} spectra of $p (\overline{p})$ in $pp$ collisions at $\sqrt{s}$= 7 TeV from various \pythia~8 modes (Simple, Vincia and Dire). Pythia~8 Simple with default value of $p_\mathrm{T}HatMin$ does not reproduce the experimental results, and the disagreement is clear over the entire {\ppt} range. On the other hand, the tuned values of $p_\mathrm{T}HatMin$ for each mode of \pythia~8 reasonably describe the experimental results for $p (\overline{p})$ particularly at {\ppt} $> 0.6$ GeV/$c$. The low {\ppt} part is not very well described by these values and the deviation is $\approx$ 20\% at {\ppt} $< 0.6$ GeV/$c$ which is clear from the model-to-data ratio plot at the bottom of the figure.  

\subsection{Particle Ratios}
In this section, we present the anti-particle to particle ($\pi^-/\pi^+$, $K^-/K^+$, $\overline{p}/p$), $(K^++K^-)/(\pi^++\pi^-)$ and $(\overline{p}+p)/(\pi^++\pi^-)$ ratios as a function of {\ppt} in $pp$ collisions at $\sqrt{s}$ = 7\;TeV from various \pythia~8 modes (Simple, Vincia and Dire). 

Figure~\ref{fig4} shows the {\ppt} dependence of various anti-particle-to-particle $\pi^-/\pi^+$, $K^-/K^+$ and $\overline{p}/p$ ratios in $pp$ collisions at $\sqrt{s}$= 7 TeV from various \pythia~8 modes (Simple, Vincia and Dire). These results are compared with the corresponding experimental data. The analysis suggests that \pythia~8 tunes, with various $p_\mathrm{T}HatMin$ values explored in this study, exhibit a reasonable capability in reproducing the measured pion $\pi^-/\pi^+$, $K^-/K^+$ and $\overline{p}/p$ yield ratios across the entire {\ppt} range. The \pythia~8  Simple tune, with the default $p_\mathrm{T}HatMin$ value, exhibits a deviation in the $K^-/K^+$ ratio at high {\ppt} bins, specifically for {\ppt} values exceeding 1.2 GeV/$c$. This deviation could potentially be attributed to statistical fluctuations in the model. In contrast, all other \pythia~8 tunes with different $p_\mathrm{T}HatMin$ values demonstrate good agreement with the experimental data for the $K^-/K^+$ ratio across the {\ppt} range. The ratios of the yields for oppositely charged particles are close to one indicates that phenomenon of pair-production at mid-rapidity. The ratios are consistent with unity across the measured {\ppt} range and hence independent from {\ppt}.

\begin{figure*}[!ht]{\label{fig4}} 
    \centering
    \includegraphics[width=0.75\textwidth]{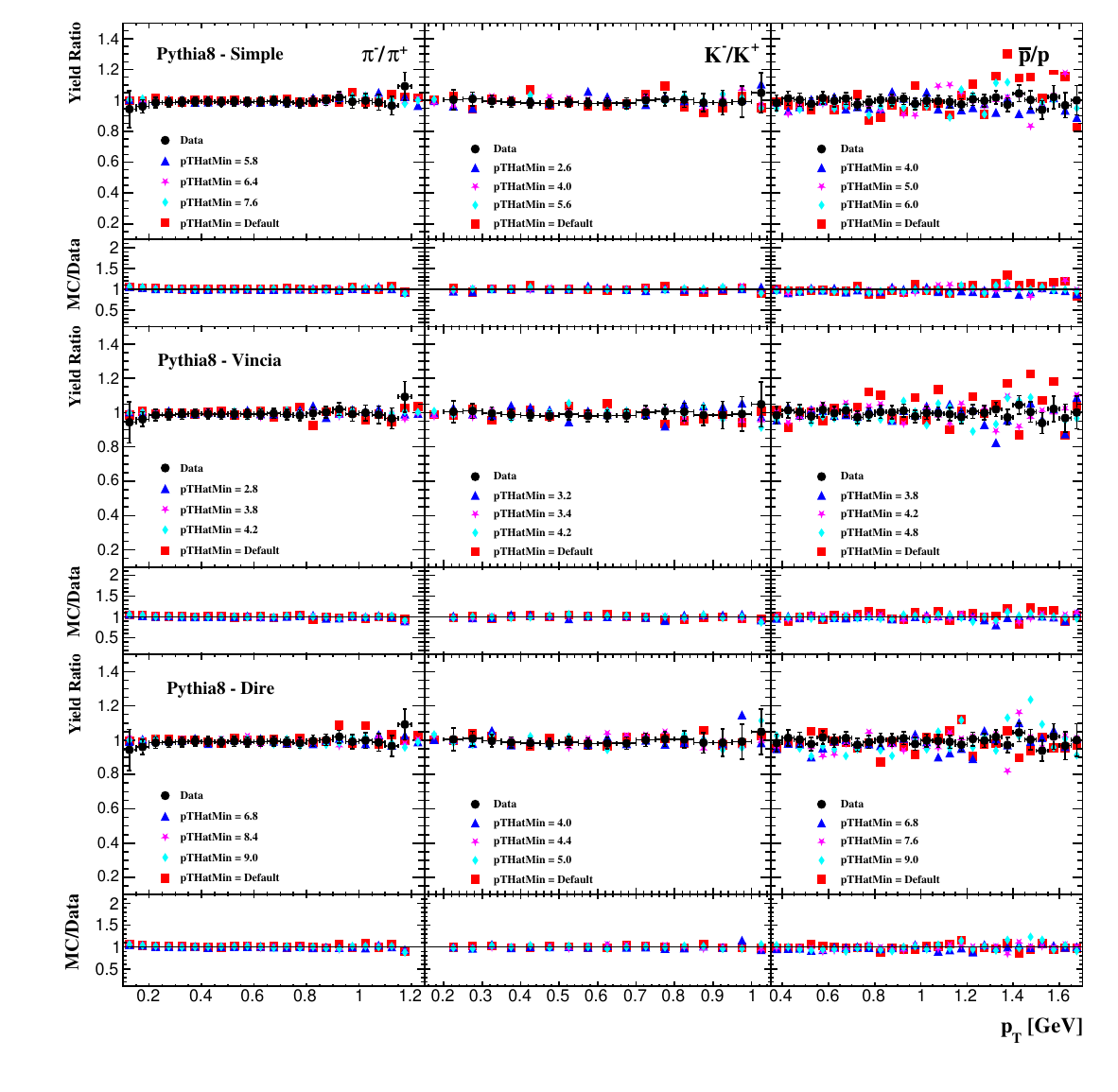}
    \caption{$\pi^-/\pi^+$, $K^-/K^+$ and $\overline{p}/p$ ratios in $pp$ collisions at $\sqrt{s}$= 7\;TeV from \pythia~8 Simple (top ), Vincia (middle) and Dire (bottom). The experimental data for comparison are taken from Ref.\cite{27}. The Model/data ratio is shown in the bottom panel.}
    \label{fig4}
\end{figure*}

Figure~\ref{fig5} shows the {\ppt}-differential $(K^++K^-)/(\pi^++\pi^-)$ (top) and $(\overline{p}+p)/(\pi^++\pi^-)$ (bottom) in $pp$ collisions at $\sqrt{s}$= 7 TeV from different modes of \pythia~8. The $(K^++K^-)/(\pi^++\pi^-)$ ratios as a function of {\ppt} are often referred to as the measure of strangeness enhancement. It can be seen from fig.~\ref{fig5} (top) that in data with increasing {\ppt}, the relative number of kaons also increases, which is expected from the simple kinematic effects and all modes of \pythia~8 underestimate that rate of growth especially at higher {\ppt} bins. This trend is also observed by the other experiments at different energies~\cite{27}. The observed enhancement in strangeness production challenges the prevailing  assumption that $pp$ collisions at these high energies can be fully explained by a simple picture of independent parton-parton scatterings.

Figure~\ref{fig5} (bottom) shows the {\ppt}-differential $(\overline{p}+p)/(\pi^++\pi^-)$ in $pp$ collisions at $\sqrt{s}$= 7 TeV from different modes of \pythia~8. This ratio serves as an indicator of the relative production
of baryons compared to mesons. It is observed that the tuned $p_\mathrm{T}HatMin$ for \pythia~8 (Simple and Dire) reasonably describe the experimental data at {\ppt} $> 0.6$ GeV/$c$ while underestimate at lower {\ppt} bins. On the other hand, all tuned values of \pythia~8 Vincia slightly overestimate the experimental data. Similar to LHCb tune of \pythia~\cite{40}, which was explicitly tuned to study the production of a wide range of particle species, the \pythia~8 at the proposed values of $p_\mathrm{T}HatMin$ reproduces the relative production of protons and pions. \pythia~8 cannot give a simultaneous and quantitative description of both the strangeness enhancement and collectivity for in $pp$ collisions. These models might require incorporating additional final-state effects, such as color ropes or junctions, to accurately describe the strangeness enhancement.

\begin{figure*}[!ht]{\label{fig5}} 
    \centering
    \includegraphics[width=0.75\textwidth]{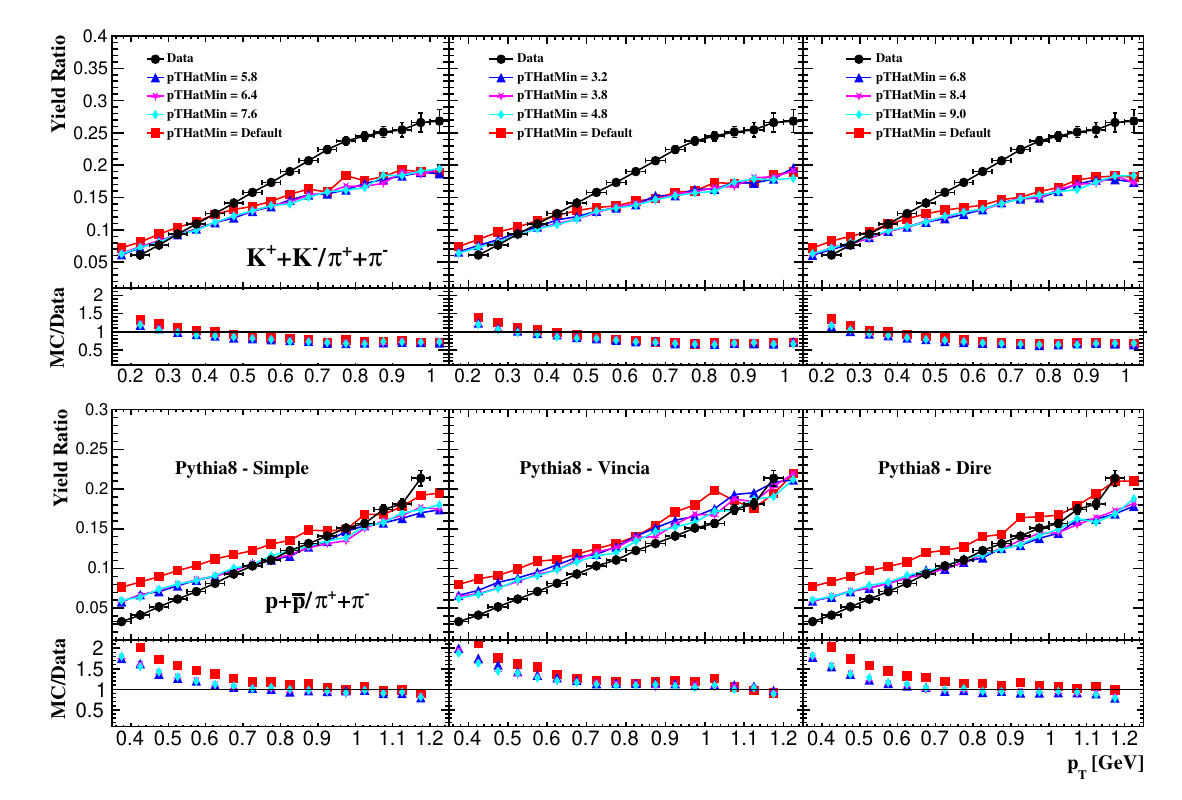}
    \caption{$(K^++K^-)/(\pi^++\pi^-)$ (top) and $(\overline{p}+p)/(\pi^++\pi^-)$ (bottom) in $pp$ collisions at $\sqrt{s}$= 7 TeV from \pythia~8 Simple (left ), Vincia (middle) and Dire (right). The experimental data for comparison are taken from Ref.\cite{27}. The Model/data ratio is shown in the bottom panel.}
    \label{fig5}
\end{figure*}

\begin{table*}[hbt!] {\label{table1}}
\scriptsize{
\caption{$p_\mathrm{T}HatMin$ values for different particles}
\vspace{-.50cm}
\begin{center}
\begin{tabular}{p{2cm}p{2cm}p{2cm}p{2cm}}\\ \hline\hline
     Particle & Simple &  Vincia& Dire  \\\hline
    $\pi^\pm$, $\pi^-/\pi^+$ &  5.8, 6.4, 7.6 & 2.8, 3.8, 4.8 & 6.8, 8.4, 9.0\\
    \hline
    $K^\pm$, $K^-/K^+$ & 2.6, 4.0, 5.6 &  3.2, 3.4, 4.2 & 4.0, 4.4, 5.0\\
    \hline
    $p(\bar p)$, $\overline{p}/p$ &  4.0, 5.0, 6.0 & 3.8, 4.2, 4.8 & 6.8, 7.6, 9.0\\
    \hline
    \hline
\end{tabular}
\label{table1}
\end{center}} 
\end{table*}
\subsection{Mean transverse momentum}
The average transverse momentum $\langle p_\mathrm{T} \rangle$ is shown as a function of mass of particle species in Fig.~\ref{fig6}. It is observed that \pythia~8 Simple with default value of $p_\mathrm{T}HatMin$ slightly underpredict the experimental data for all particle species, however, the trend is similar. For other values of $p_\mathrm{T}HatMin$, the \pythia~8 well describe the $\langle p_\mathrm{T} \rangle$ of pions and kaons. On the other hand, none of the tunes provides an acceptable description of the mass dependence of $\langle p_\mathrm{T} \rangle$ for protons and the observed values from \pythia~8 largely underestimate the experimental measurements.  
\pythia~8 shows increasing trend with increase in energy similar to the data. This observation is interesting because the momentum spectra themselves suggest an increasing contribution from hard scattering processes, which typically lead to higher $\langle p_\mathrm{T} \rangle$ particles

\begin{figure*}[!ht]{\label{fig6}} 
    \centering
    \includegraphics[width=\textwidth]{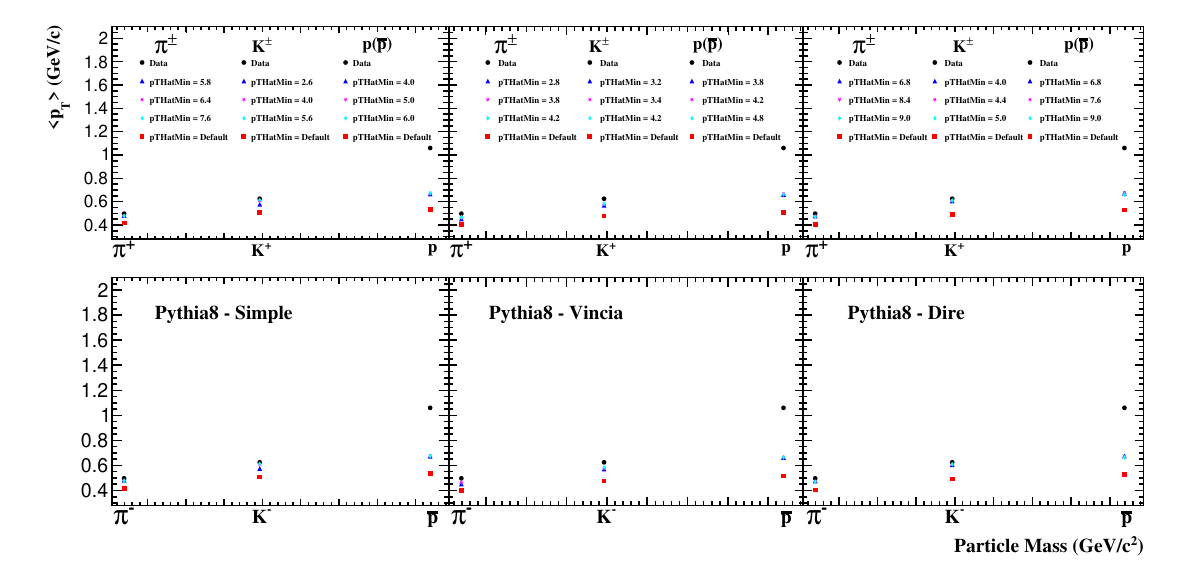}
    \caption{The average transverse momentum $\langle p_\mathrm{T} \rangle$ for $\pi^+, K^+, p$ (top) and for $\pi^-, K^-, \overline{p}$ (bottom) as a function of mass of particle species in $pp$ collisions at $\sqrt{s}$= 7 TeV from \pythia~8 Simple (left ), Vincia (middle) and Dire (right). The experimental data for comparison are taken from Ref.\cite{27}.}
    \label{fig6}
\end{figure*}

\section{Conclusion}\label{sec4}
In this study, we reported a comprehensive analysis of various observables, including transverse momentum ({\ppt}) spectra, particle ratios, and mean transverse momentum of identified particles ($\pi^\pm$, $K^\pm$ and $p(\bar{p}$) in $pp$ collisions at $\sqrt{s}$ = 7 TeV using different modes of \pythia~8 Monte- Carlo event generator. Our analysis focused on understanding the influence of the $p_\mathrm{T}HatMin$ parameter on these observables to better reproduce experimental data from the CMS collaboration and assessing the performance of different \pythia~8 modes (Simple, Vincia, and Dire). We found that the default $p_\mathrm{THatMin}$ values in \pythia~8 failed to reproduce experimental data. However, fine tuning of $p_\mathrm{THatMin}$ significantly improved agreement, particularly for $\pi^\pm$ spectra across all \pythia~8 modes. \pythia~8 Dire provides better agreement for $K^\pm$ spectra at higher {\ppt}, while Simple and Vincia better describe the data at lower {\ppt}. Challenges remained in accurately reproducing $p(\bar{p})$ spectra, especially at lower {\ppt}. Furthermore, particle ratios revealed consistent discrepancies, notably in $(K^++K^-)/(\pi^++\pi^-)$ ratios at higher {\ppt} across all modes. The strangeness enhancement challenges the prevailing  assumption that $pp$ collisions at these high energies can be fully explained by a simple picture of independent parton-parton scatterings suggesting the need for models that incorporate additional final-state effects like color ropes or junctions. Similarly, $(\overline{p}+p)/(\pi^++\pi^-)$ ratios exhibited deviations at lower {\ppt}, with Simple and Dire modes underestimating, and Vincia mode slightly overestimating experimental observations. Discrepancies in $\langle p_\mathrm{T} \rangle$ underscored the nuanced dependencies on particle species. \pythia~8 cannot give a simultaneous and quantitative description of both the strangeness enhancement and collectivity for in $pp$ collisions. These models might require incorporating additional final-state effects, such as color ropes or junctions, to accurately describe the strangeness enhancement.


\end{multicols}{}
\end{document}